\documentclass[sigconf]{acmart}

\usepackage{lineno,hyperref}
\usepackage{booktabs} % For formal tables
\usepackage{color}
\usepackage{multirow}
\usepackage{array}
\usepackage{ragged2e}
\usepackage{float} % para usar [H]
\usepackage{balance}
\usepackage[latin1]{inputenc}
\usepackage{natbib} 
\usepackage{setspace}
\usepackage{booktabs}
\usepackage{soul}
\usepackage{placeins}
\usepackage{framed}
\usepackage{url}
\usepackage[normalem]{ulem}
\usepackage{caption} 
\usepackage[inline]{enumitem}
\usepackage{caption}
\usepackage{subcaption}
\usepackage{amsmath}
\usepackage{rotating}
\usepackage{graphicx}
\usepackage{amsmath}
\usepackage{amssymb,amsfonts}
\usepackage{algorithmic}
\usepackage{scrtime}
\usepackage{graphicx}
\usepackage{textcomp}
\usepackage[latin1]{inputenc}
\usepackage{url}
\usepackage{xcolor}
\usepackage{colortbl}

\usepackage{graphicx} % figuras
 \usepackage{float}
\usepackage[inline]{enumitem}
\usepackage{framed}
\usepackage{tabularx}
\usepackage{soul}
\usepackage{lipsum}
\usepackage{setspace}
\usepackage{soul}
\usepackage{ragged2e}
\usepackage{float} % para usar [H]
\usepackage{booktabs} % To thicken table lines
\usepackage{amssymb}
\usepackage{array}
\usepackage {multirow}

\newcolumntype{P}[1]{>{\centering\arraybackslash}p{#1}}

\modulolinenumbers[5]

\AtBeginDocument{%
  \providecommand\BibTeX{{%
    \normalfont B\kern-0.5em{\scshape i\kern-0.25em b}\kern-0.8em\TeX}}}

\setcopyright{acmcopyright}
\copyrightyear{2020}
\acmYear{2019}
\acmDOI{10.xxxxx}

\copyrightyear{2020} 
\acmYear{2020} 
\setcopyright{acmcopyright}
\acmConference[EASE 2020]{Evaluation and Assessment in Software Engineering}{April 15--17, 2020}{Trondheim, Norway}
\acmBooktitle{Evaluation and Assessment in Software Engineering (EASE 2020), April 15--17, 2020, Trondheim, Norway}
\acmPrice{15.00}
\acmDOI{10.1145/3383219.3383233}
\acmISBN{978-1-4503-7731-7/20/04}

\begin{document}

\title{Publication Bias: A Detailed Analysis of Experiments Published in ESEM}

\author{Rolando P. Reyes Ch.}
 
\affiliation{%
  \institution{Universidad de las Fuerzas Armadas ESPE}
  \streetaddress{Av. General Rumi\~nahui s/n.}
  \city{Sangolqu\'i} 
  \country{Ecuador} 
  \postcode{28660}
}
\email{rpreyes1@espe.edu.ec}

\author{Oscar Dieste}
 
\affiliation{%
  \institution{Universidad Polit\'ecnica de Madrid}
  \city{Madrid, Boadilla del Monte} 
  \country{Spain} 
  \postcode{28660}
}
\email{odieste@fi.upm.es}

\author{Efra\'in R. Fonseca C.}
 
\affiliation{%
  \institution{Universidad de las Fuerzas Armadas ESPE}
  \streetaddress{Av. General Rumi\~nahui s/n.}
  \city{Sangolqu\'i} 
  \country{Ecuador} 
  \postcode{28660}
}
\email{erfonseca@espe.edu.ec}

\author{Natalia Juristo}
 
\affiliation{%
  \institution{Universidad Polit\'ecnica de Madrid}
  \city{Madrid, Boadilla del Monte} 
  \country{Spain} 
  \postcode{28660}
}
\email{natalia@fi.upm.es}

\renewcommand{\shortauthors}{Reyes et al.}


\begin{abstract}

\textit{Background:} Publication bias is the failure to publish the results of a study based on the direction or strength of the study findings. The existence of publication bias is firmly established in areas like medical research. Recent research suggests the existence of publication bias in Software Engineering.
\textit{Aims:} Finding out whether experiments published in the International Workshop on Empirical Software Engineering and Measurement (ESEM) are affected by publication bias.
\textit{Method:} We review experiments published in ESEM. We also survey with experimental researchers to triangulate our findings.
\textit{Results:} ESEM experiments do not define hypotheses and frequently perform multiple testing. One-tailed tests have a slightly higher rate of achieving statistically significant results. We could not find other practices associated with publication bias.
\textit{Conclusions:} Our results provide a more encouraging perspective of SE research than previous research: (1) ESEM publications do not seem to be strongly affected by biases and (2) we identify some practices that could be associated with p-hacking, but it is more likely that they are related to the conduction of exploratory research.

\end{abstract}





\maketitle

\section{Introduction}\label{intro}

Publication bias is defined as \textit{failure to publish the results of a study 'on the basis of the direction or strength of the study findings'} \cite{DeVito53}. Typically, studies containing positive (i.e., direction), or statistically significant (i.e., strength) results, are more likely to be published than studies reporting negative or non-significant results \cite{Song2013open}.

Publication bias has been observed in disciplines such as medical research \cite{easterbrook1991publication,Dwan2008PLOS}, but not in others, e.g., personnel selection research \cite[pp. 493-495]{Schmidt-2015-Meta-Analysis}. In Software Engineering (SE), J\o rgensen et al. \cite{jorgensen2016incorrect} reviewed the ratio of statistically significant tests in 150 randomly selected experiments. Such a ratio (51\%) supports the existence of substantial publication bias in SE.

Publication bias promotes questionable research practices \cite[Figure 1]{munafo2017manifesto}: Failure to control for cognitive biases, analytical flexibility or data dredging (\textit{fishing} or \textit{p-hacking}) \cite{nuzzo2014statistical}. Ioannidis \cite{ioannidis2005most}, in a foundational paper, argued that most experimental results are false, pointing out almost the same reasons: little criticism of the posed research questions, excessive design and analysis flexibility, and research biases.

In a previous study \cite{Reyes2018icse}, we searched for statistical errors in experiments published in the International Conference on Software Engineering (ICSE). We detected that a significant amount of experiments do not report statistical hypothesis, do not check the assumption of the inference tests, and perform multiple testing without any correction to avoid increasing type-I errors. These practices are typically associated with publication bias.

The Int'l Workshop on Empirical Software Engineering and Measurement (ESEM) is the flagship conference on experimental SE. In this research, we review experiments published in ESEM searching for practices associated with publication bias.  We also survey with SE experimenters to triangulate the literature review results. 

The \textbf{contribution of this paper is twofold}:

\begin{itemize}

\item We provide a more encouraging image of SE research than previous research \cite{jorgensen2016incorrect,Reyes2018icse}. ICSE is the flagship SE conference, but it has a general, i.e., non-experimental, character. The same applies to the sources surveyed by J\o rgensen and colleagues. Although the current situation can be improved, ESEM publications do not seem to be strongly affected by biases.

\item We identify some practices that could be associated with \textit{p-hacking}. However, a more likely explanation is our almost complete ignorance about SE phenomena, that leads the conduction of a good deal of exploratory research.

\end{itemize}

This paper is organized as follows: Section~\ref{sec:background} introduces the practices that may suggest the existence of publication bias. Section~\ref{sec:methodology} state the research questions. Section~\ref{sec:paper_review} reports the literature review, and Section~\ref{sec:survey-experimenters} the survey to experimenters. The paper finishes with the threats to validity in Section~\ref{sec:threats-validity}, and the conclusions in Section~\ref{sec:conclusion}.
\section{Background}\label{sec:background}

Publication bias means papers containing statistically significant results have a higher likelihood of being published. Experimenters may be tempted to carry out some questionable practices that increase the chances to achieve statistical significance. According to {Munaf{\`o} et al. \cite[Figure 1]{munafo2017manifesto}, these practices are\footnote{Figure 1 in \cite{munafo2017manifesto}, and the caption of the same figure, are not totally consistent. We provide our personal interpretation herein.}:

\subsection{Failure to control for bias}

Once experimental data is available, data visualization and exploratory analyses are usual. These practices cannot be questioned in principle, but they convey a risk: researchers may perceive patterns or regularities that suggest relationships between variables. Inadvertently (or not), these relationships can find their way into the study as genuine hypotheses. 

HARKing (Hypothesizing After Results are Known) \cite{kerr1998harking} means that \textit{post hoc} hypotheses are presented and analyzed as they were \textit{a priori} hypotheses. When the patterns/regularities are strong enough, \textit{post hoc} hypotheses yield statistically significant results independently of any other consideration, e.g., statistical power.

HARKing can adopt different forms \cite[pp. 197-198]{kerr1998harking}. When \textit{post hoc} hypotheses are presented as \textit{a priori} hypotheses, HARKing cannot be detected. However, some other practices indicate (but do not confirm) that HARKING is present:

\begin{itemize}

\item The experiment does not contain hypotheses, but inference tests are applied to the data. Although these tests can be based on non-reported \textit{a priori} hypotheses, such tests may be conducted on an opportunistic or exploratory manner.

\item The experiment contains, besides the experimental hypothesis, additional hypotheses. They typically explore relationships between experimental and non-experimental (e.g., demographics) data. These hypotheses are usually termed \textit{post hoc} hypotheses in the experimental literature.

\end{itemize}

\subsection{Analytical flexibility}

Experimenters have complete control of the data acquisition process and analysis procedures. Harmless decisions, e.g., outlier removal, dataset reduction, or the introduction of controlled variables (gender, experience), can influence the statistical significance of the tests\footnote{These practices can also be seen as instances of data dredging, e.g., see \cite[pp. 169]{indrayan2016concise}. Different authors use the same terms with (slightly) different meanings. We provide here a coherent but necessarily partial picture.}. Again, as in the case of the HARKing, it is generally impossible to assess the researchers' intention because it is not reflected in the written reports. 

One exception is the choice of inferential tests. For most designs, simple tests suffice, such as ANOVA, or the corresponding non-parametric counterpart (Kruskal-Wallis). When unusual, sophisticated tests are used, that is an indication (again, not a confirmation) of \textit{p-hacking} \cite[pp. 147-150]{Vickers2010}.

\subsection{Data dredging}

Data dredging is the practice of performing comparisons within a dataset to achieve statistically significant results \cite[pp. 169]{indrayan2016concise}. This practice is favored when a large number of independent and, more frequently, dependent variables, are used in a study. Data dredging can take several forms:

\begin{itemize}

\item Performing multiple comparisons, beyond what is required to test the statistical hypotheses.

\item Posing multiple hypotheses \footnote{Multiple hypotheses are also an indication of HARKing, but it cannot be assessed in written reports.} is also an indication (again, not a proof) of data dredging. In most cases, these hypotheses test the relationship between a single independent variable and multiple dependent variables. In practice, they are equivalent to multiple testing.

\item Making the analysis more powerful. There are several possible procedures (e.g., switching tests), but none of them guarantee that the power increases. One exception is switching the tails of the tests, from 2-tailed to 1-tailed. 2-tailed tests are inherently less powerful. 1-tailed tests are sometimes used not because existing knowledge suggests a directional effect, but to increase the chances of achieving statistically significant results.

\end{itemize}

\section{Research questions and methodology}\label{sec:methodology}

Publication bias seems to affect SE \cite{jorgensen2016incorrect} , as well as other more mature disciplines \cite{easterbrook1991publication,Dwan2008PLOS}. However, experimental research in SE has not achieved standardization. At least in principle, experimental research may have different levels of rigor in each particular community. ESEM aims to be the ''the premier conference for presenting research results related to empirical software engineering'' \cite{esem2019}. Thus, we wonder:

\begin{framed}
\noindent\textbf{RQ1: Is there evidence of publication bias at ESEM?}
\end{framed}

To answer this question, we have conducted a literature review of experiments published in ESEM between 2007 and 2016 (10 years in total), seeking signs of failure to control for cognitive biases, analytical flexibility, and data dredging.

The literature review is reported in Section~\ref{sec:paper_review}. Unfortunately, most of the evidence is ambiguous. For instance, the lack of an explicit hypothesis definition does not automatically imply HARKing. Likewise, the switch of a 2-tailed test into a 1-tailed one suggests a \textit{p-hacking}, but we cannot rule out alternative explanations, e.g., a mistake overlooked by authors and reviewers. Therefore, we propose another research question:

\begin{framed}
\noindent\textbf{RQ2: Why do researchers carry out questionable practices?}
\end{framed}

To answer this question, we have conducted a survey with SE experimenters (see Section~\ref{sec:survey-experimenters}). Their answers contextualize the literature review and triangulate our findings.
\section{Literature review}\label{sec:paper_review}

To answer RQ1, we have conducted a systematic literature review according to Kitchenham et al. \cite{Kitchenham2007-SystematicLitReview}. The following sections describe the review objectives, design, execution, results, and the main findings obtained from the review.

\subsection{Review objectives}\label{sec:paper_review:objectives}

The literature review aims to identify whether experiments published at ESEM carry out practices (described in Section~\ref{sec:background}) that suggest publication bias. More concretely, we propose the following review objectives:

\begin{itemize}

\item \textbf{Failure to control for bias}: 

  \begin{enumerate}

  \item Are hypotheses explicitly defined?

  \item Are there \textit{post hoc} tests?

  \end{enumerate}
   
\item \textbf{Analysis flexibility}:
   
  \begin{enumerate}\setcounter{enumi}{2}

  \item Which analysis procedures do ESEM experiments apply?

  \end{enumerate}
   
\item \textbf{Data dredging}:
   
  \begin{enumerate}\setcounter{enumi}{3}

  \item How many explicit hypotheses are defined per experiment?
   
  \item How many tests are conducted per experiment?

%  \item How many tests are conducted per hypothesis?

  \item What is the ratio of statistically significant tests?

  \item Does the hypothesis tail match the analysis tail?
   
  \item Are 1-tailed tests used to increase power?

  \end{enumerate}

\end{itemize}

\subsection{Inclusion and exclusion criteria}

Papers qualify as experimental when they have:

\begin{itemize}

\item \textit{Group assignment}: experimental units are assigned (randomly or not) to groups.

\item \textit{Comparative goal}: they aim to compare some response variables across groups.

\item \textit{Inference}: (frequentist) statistical tests are used to reveal differences among groups. 

\item \textit{Experimental data}: The data is generated as a result of the experimental manipulations; previously existing data is not used.

\end{itemize}

The 1\textsuperscript{st} inclusion criterion asserts the paper's experimental (or quasi-experimental) character \cite{Shadish2002}. The 2\textsuperscript{nd} and 3\textsuperscript{rd} guarantee the use of hypothesis testing. The 3\textsuperscript{rd} criterion would have excluded experiments analyzed under the Bayesian framework if there were any; none of the reviewed papers applied Bayesian statistics.

A large number of papers use existing datasets, e.g., PROMISE repository \cite{Sayyad-Shirabad+Menzies:2005}, data extracted from open source repositories, etc. The studies that rely on these data do not have an experimental character, i.e., the data is not obtained as the result of some experimental manipulation. As the data is pre-existing (either readily available, as the case of the PROMISE repository, or available to be processed, as the case of open-source repositories), such studies can be properly termed as observational studies. The 4\textsuperscript{th} criterion makes a distinction between experimental (or quasi-experimental) and observational studies. The latter are excluded from our literature review. Such exclusion does not lead to negative results. In fact, observational studies depart widely from usual experimental standards; their inclusion would have made the review results considerably worse.

\subsection{Execution}

We reviewed experimental papers published in ESEM between 2007-2016. Two researchers (R. Reyes \& O. Dieste) screened\footnote{Details are available at \url{https://goo.gl/PSjjQu}.} the titles, abstracts, and keywords of all papers independently, using the ACM and IEEE digital libraries. Discrepancies were solved by consensus. 

Fleiss' $\kappa=0.62$, representing substantial agreement \cite{fleiss2013statistical}. In total, 387 papers were screened and 55 papers initially selected. After a detailed review, 6 papers were removed because they did not meet the inclusion criteria. Finally, the primary study set was composed of 49 experiments.

The first author (R. Reyes) created an extraction form\footnote{The form and the raw data are available at \url{https://goo.gl/PSjjQu}.}, which stems from the review objectives described in section \ref{sec:paper_review:objectives}. Two researchers (O. Dieste \& R. Fonseca) piloted the form, suggesting several changes and general improvements to the wording. Two researchers (R. Reyes \& O. Dieste) extracted the information from the primary studies independently. Both datasets were later compared and corrective measures were taken in case of disagreement.

\subsection{Review results}

The collected data were summarized using 2-way tables and tree-like representations. The results are described below.

\subsubsection{Are hypotheses explicitly defined?}

Shortly after commencing the review, we realized that the primary studies define hypotheses at different levels of abstraction:

\begin{itemize}

\item Research goal or aim: A high-level statement describing the purpose of the experiment, e.g.:

  \begin{quote}
  \textit{We wanted to find out whether CFT are less error-prone and whether the participants would favor CFT over FT} \cite{jung2013experimental}.
  \end{quote}

\item Research hypothesis: A declaration of the relationship between independent and dependent variables that drives the research design \cite{mcpherson2001teaching}. For instance:

  \begin{quote}
  \textit{$H_{21}$: When using CFT, consistency between the system description and the safety analysis model is perceived differently than when using FT.} \cite{jung2013experimental}
  \end{quote}

\item Research questions: A research goal breaks down into several research hypothesis rather frequently. Although unnecessary from a technical viewpoint, researchers tend to create \textit{research questions} to report this refinement process explicitly. For instance, the research questions below appear also in \cite{jung2013experimental}:

  \begin{quote}

  \textit{$RQ_1$: Will the application of the CFT yield the same quality of the resulting safety model as a model built with FT?}

  \textit{$RQ_2$: Is CFT perceived differently than FT with regard to consistency, clarity, and maintainability?} 

  \end{quote}

\item Statistical hypothesis: Finally, research hypotheses can adopt an analytic formulation, using the usual null/alternate format. The statistical hypothesis contains the keys elements (estimators, tails, etc.) that drive statistical analysis \cite{mcpherson2001teaching}. Again using \cite{jung2013experimental} as example:

  \begin{quote}

  \textit{$H0_{21}$: $\mu CFT = \mu FT$}

  \textit{$H_{21}$: $\mu CFT \neq \mu FT$}

  \end{quote}

\end{itemize}

The previous elements (research goal/question/hypothesis, and statistical hypothesis) do not usually appear together in the same experiment. Jung et al. \cite{jung2013experimental} is one of the few cases in which the four are explicitly reported. In turn, it is rather common that one or several are missing (in particular, the research hypotheses and the statistical hypotheses). The same problem has been observed in other areas \cite{cho2013two}.

Fig.~\ref{fig:RQ11} summarizes how ESEM experiments define hypotheses. Almost all experiments (96\%) specify the research goal. However, only 55\% of them contain research hypotheses and, looking further down, the declaration of statistical hypotheses (29\%) decreases considerably. 

The tails of the tests are defined in a larger proportion (49\%) than the statistical hypotheses. The reason is that researchers specify the type of tail in the research hypothesis rather frequently. The leftmost branch corresponds with the orthodox use of hypothesis testing; only 12\% of the experiments conform to it. 

\begin{figure*}[htbp]
\centering
\includegraphics[width=0.65\textwidth]{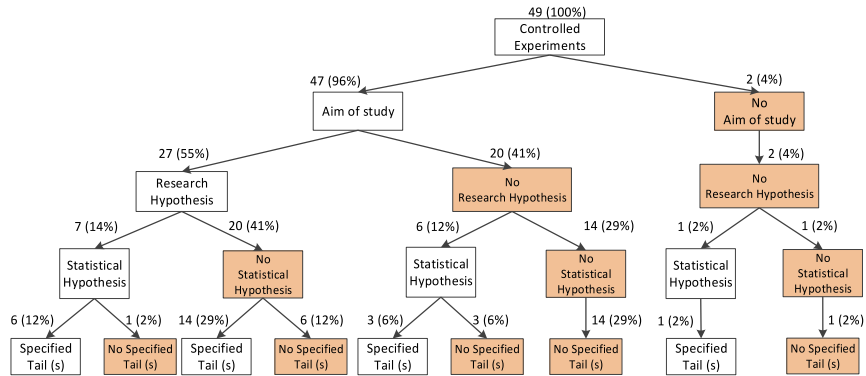}
\caption{Relationships among aims, research hypotheses, statistical hypotheses and tails} \label{fig:RQ11}
\end{figure*}

27\% of the experiments do not include research or statistical hypotheses. This is rather unusual since a declaration of purpose (which variables are being examined, and why) predates experimental operation, not to mention it goes against the recommendations of standard experimentation textbooks and guidelines, e.g.: Juristo \& Moreno explicitly mention research hypotheses \cite[pp. 49]{juristo2013basics}; both Wohlin et al. and Jedlitschka et al. emphasize the usage of statistical hypotheses  \cite{wohlin2012experimentation,Jedlitschka2005-GuideLinesESE}. The underlying reason seems to be the widespread usage of research questions instead of hypotheses. Table~\ref{tab:RQ_vs_research} shows that 70\% of the papers contain RQs but not research hypotheses. Table~\ref{tab:RQ_vs_statistical} reports the same figure\footnote{The values are alike by chance; it is not a mistake.} (70\%) for the statistical hypotheses.

\begin{table}[htbp]
\scriptsize
\centering
\caption{RQs vs. Research Hypotheses Crosstab}
\label{tab:RQ_vs_research}
\begin{tabular}{c c c c}
\cline{2-4}
& \begin{tabular}[c]{@{}c@{}} \textbf{Have research} \\ \textbf{ hypothesis}\end{tabular} 
& \begin{tabular}[c]{@{}c@{}}\textbf{Do not have} \\ \textbf{research hypothesis}\end{tabular} 
& \begin{tabular}[c]{@{}c@{}}\textbf{Total}\end{tabular} 
\\ \hline \hline
\multicolumn{1}{c}{\textbf{Have RQs}} & 14 (29\%)                                                                & 20 (41\%) & \textbf{34 (70\%)}                                                                \\
\multicolumn{1}{c}{\textbf{Do not have RQs}} & 13 (26\%)                                                                  & 2 (4\%) & \textbf{15 (30\%)}                                                                \\ \hline
\multicolumn{1}{c}{\textbf{Total}} & \textbf{27 (55\%)}                                                                  & \textbf{22 (45\%)}                                                                 \\ \hline
\end{tabular}
\end{table}

\begin{table}[htbp]
\scriptsize
\centering
\caption{RQs vs. statistical hypotheses crosstab}
\label{tab:RQ_vs_statistical}
\begin{tabular}{c c c c}
\cline{2-4}
& \begin{tabular}[c]{@{}c@{}} \textbf{Have statistical} \\ \textbf{ hypothesis}\end{tabular} 
& \begin{tabular}[c]{@{}c@{}}\textbf{Do not have} \\ \textbf{statistical hypothesis}\end{tabular} 
& \begin{tabular}[c]{@{}c@{}}\textbf{Total}\end{tabular} 
\\ \hline \hline
\multicolumn{1}{c}{\textbf{Have RQs}} & 10 (21\%)& 24 (49\%) & \textbf{34 (70\%)}\\
\multicolumn{1}{c}{\textbf{Do not have RQs}} & 4 (8\%) & 11 (22\%) & \textbf{15 (30\%)}\\ \hline
\multicolumn{1}{c}{\textbf{Total}} & \textbf{14 (29\%)} & \textbf{35 (71\%)}\\ \hline
\end{tabular}
\end{table}

\subsubsection{Are there \textit{post hoc} tests?}

Only $\frac{15}{49} = 31\%$ of the experiments contain \textit{post hoc} tests. \textit{Post hoc} tests were conducted after the inference tests. In 12 out of 15 cases (80\% of the total), inference tests already yielded statistically significant results, so \textit{post hoc} tests were unnecessary (from the publication bias perspective).

The \textit{post hoc} tests performed in each experiment have their own peculiar characteristics, but they can be roughly classified into three groups:

\begin{itemize}

\item Subjects are decomposed into subgroups on the basis of some characteristic, e.g., experience, and the hypotheses are re-tested.

\item During hypothesis testing, some independent variables that have a role in the experimental design were not examined. They are examined during \textit{post hoc} testing.

\item Correlations (usually bivariate) are run between independent and dependent variables.

\end{itemize}

\subsubsection{Which analysis procedures do ESEM experiments apply?}

Fig.~\ref{fig:12tailed_coherence} shows (besides other aspects that we will not discuss in this moment) the types of tests used in ESEM experiments. 

None of the tests is uncommon; actually, they constitute a rather basic statistical toolset, e.g., t-test, Mann-Whitney, Wilcoxon, ANOVA, Kruskal-Wallis, etc. The \textit{Friedman} and \textit{McNemar} tests (a non-parametric repeated-measures and an alternative to the Fisher exact test, respectively) are somehow unusual, but by no means ''sophisticated''.

\begin{figure*}[htbp]
\centering
\includegraphics[width=0.65\textwidth]{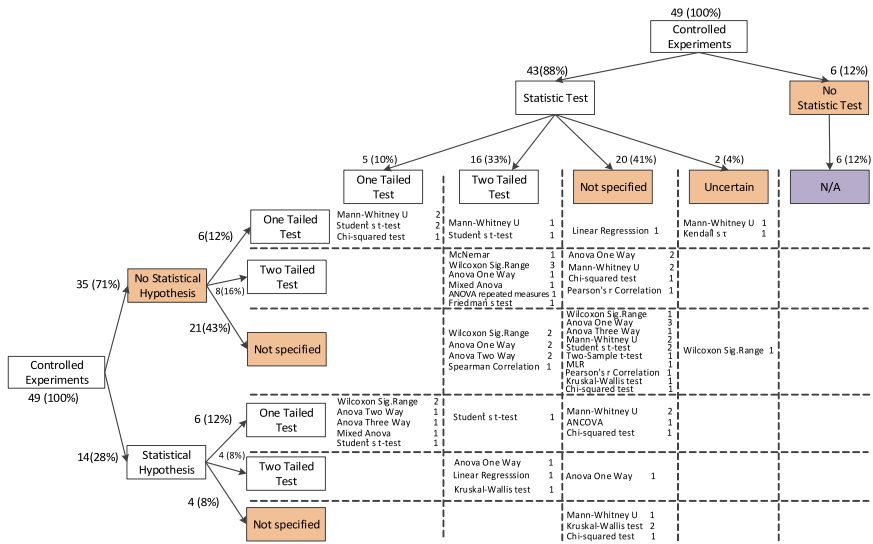}
\caption{Relationship between statistical hypotheses and tests} \label{fig:12tailed_coherence}
\end{figure*}

\subsubsection{How many explicit hypotheses are defined per experiment?}

Fig.~\ref{fig:hypothesisbyexperiment} shows a histogram. The x-axis represents the number of hypotheses defined per experiment, and the y-axis the number of experiments of each kind. Most of the experiments declare $\leq 2$ hypotheses (2.41 on average). Experiments with more than 4 hypotheses are rare.

\begin{figure}[htbp]
\centering
\includegraphics[scale=0.25]{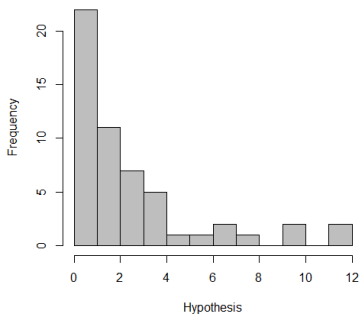}
\caption{Number of hypotheses are defined per experiment} \label{fig:hypothesisbyexperiment}
\end{figure}

\subsubsection{How many tests are conducted per experiment?}

There is a large difference between the number of hypotheses per experiment and the number of tests actually conducted. As shown in Fig.~\ref{fig:statisticaltestbyexperiment}, most of the experiments conduct $\leq 12$ tests. The average is 7.95, roughly 3 times the number of hypotheses.

The reason for the difference is that research hypotheses typically make reference to the main factor of interest, but not to other independent variables of the design, e.g., task, type of object, etc. In some cases, e.g., non-normality, the independent variables are not analyzed jointly, e.g., using ANOVA, but they are tested in sequence using non-parametric tests. This increases the number of tests as compared to the number of hypotheses.

\begin{figure}[htbp]
\centering
\includegraphics[scale=0.25]{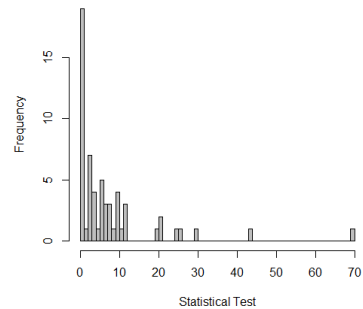}
\caption{How many tests are conducted per experiment?} \label{fig:statisticaltestbyexperiment}
\end{figure}

%\subsubsection{How many tests are conducted per hypothesis?}
%
%Max:44, Min:0, Avg: 3.08
%
%\begin{figure}[htbp]
%\centering
%\includegraphics[scale=0.3]{eps/statisticaltestbyhypothesis}
%\caption{Number of tests are conducted per hypothesis} \label{fig:statisticaltestbyhypothesis}
%\end{figure}

\subsubsection{What is the ratio of statistically significant tests?}

On average, an experiment published in ESEM contains 3.87 statistically significant tests on average. However, there is a large variation at the level of individual experiments. As shown in Fig.~\ref{fig:significantestbyexperiment}, roughly half of the papers have $\leq 1$ significant tests only. Overall, there are 179 statistically significant tests out of 480 conducted tests. This gives a significant test ratio of $\frac{179}{480} = 0.37$.

\begin{figure}[htbp]
\centering
\includegraphics[scale=0.25]{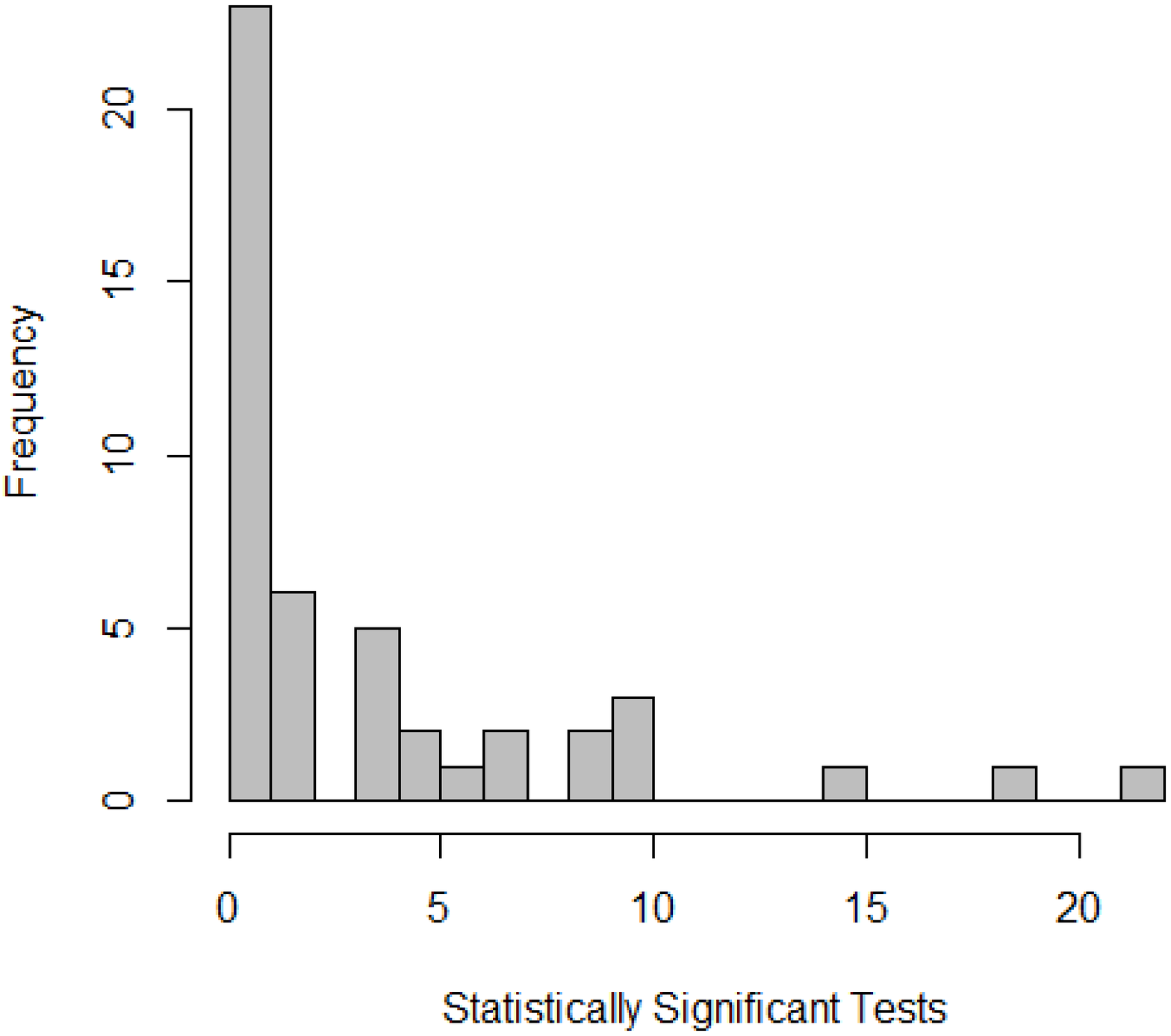}
\caption{How many statistically significant tests are conducted per experiment?} \label{fig:significantestbyexperiment}
\end{figure}

\subsubsection{Does the hypothesis tail match the analysis tail?}

In addition to the statistical tests used in ESEM experiments, Fig.~\ref{fig:12tailed_coherence} shows the tails associated with those tests. Each test is related to two tails: the one defined in the corresponding research/statistical hypothesis and the actual tail employed in the analysis.

In almost all cases, the tail of the hypothesis and the tail of the test match. Only in three cases, 1-tailed hypotheses were tested using 2-tailed tests. The most likely cause is a mistake in the definition of the hypotheses' tails. Switching from a 1-tailed to a 2-tailed test does not increase power, but actually decreases it. Those mistakes cannot be thus associated with research bias.

A challenge to this interpretation comes from the fact that, in a large number of cases, the definition of the tail is missing either in the research/statistical hypothesis and/or the test. In our opinion, the lack of definition of the tests cannot be understood as evidence of publication bias. The vast majority of tests specifying the type of tail are 2-tailed; there is not any reason to think that \textit{all} the other tests are 1-tailed. In this situation, concealing the information about the tails would have had no effect in terms of power. 

Although not related to publication bias, we wish to point out in passing some inconsistencies in Fig.~\ref{fig:12tailed_coherence}. ANOVA's are inherently 2-tailed. Even so, they have been used as 1-tailed tests in four cases. Likewise, Chi-squared tests have been used in the primary studies to test contingency tables exclusively. Contingency tables are inherently 2-tailed, but in one case the authors claim that the test is 1-tailed. Similar problems could have happened in other tests whose tails are not explicitly declared. In any case, these problems do not seem to be associated with publication bias, but to a shallow spread of statistical knowledge in the SE community.

% Esto ya no es cierto, pero lo dejo por si acaso
%
% However, coherence is achieved at a cost: as indicated above, the vast majority of experiments miss the specification of tails, both in the hypotheses \textit{and} tests sides. En otras palabras, puede haber muchos cambios en el tipo de cola que no podemos observar. On this regard, an interesting pattern can be observed in Fig.~\ref{fig:12tailed_coherence}: ``complex'' tests, e.g. ANOVA and Kruskal-Wallis, with only one exception, appear in the ``not specified'' group. These tests are inherently 2-tailed, but teaching and reference books do not emphasize this fact. Furthermore, ANOVA and Kruskal-Wallis are used when factors contain more than two levels. Statistical hypotheses with more than two levels are probably perceived as contrived and not explicitly stated in the manuscripts.

\subsubsection{Are 1-tailed tests used to increase power?}

Table~\ref{Tab:12TailedCase} shows the relationship between the types of tails of the statistical tests (notice that not all papers define the tails) and the results of the corresponding inference tests. When 1-tailed tests are conducted, the ratio of statistically significant tests is $\frac{51}{51+26} = 66$\%. In the case of 2-tailed tests, the ratio is $\frac{61}{61+74} = 45$\%. The obvious conclusion is that 1-tailed tests are associated with higher rates of statistically significant tests.

% \usepackage{multirow}
\begin{table}[htbp]
\scriptsize
\centering
\caption{Relationship between the tail type and the significance of statistical tests}
\label{Tab:12TailedCase}
\begin{tabular}{llccc}
\cline{3-5}
                                                                      & \multicolumn{1}{c}{}  & \multicolumn{3}{l}{\textbf{Results by Total Hypothesis}}                                                             \\ \hline
\multicolumn{1}{l}{}                                                & \multicolumn{1}{l}{}  & \multicolumn{1}{l}{\textbf{Sig.}} & \multicolumn{1}{l}{\textbf{No. Sig.}} & \multicolumn{1}{l}{\textbf{Uncertain}} \\ \hline \hline
\multicolumn{1}{l|}{\multirow{5}{*}{\textbf{Type of tail selected}}} & \textbf{1-Tailed}      & 51                                 & 26                                     & 0                                       \\
\multicolumn{1}{l|}{}                                                & \textbf{2-Tailed}      & 61                                 & 74                                     & 5                                       \\
\multicolumn{1}{l|}{}                                                & \textbf{Not specified} & 53                                 & 142                                    & 2                                       \\
\multicolumn{1}{l|}{}                                                & \textbf{N/A}           & 1                                  & 23                                     & 5                                       \\
\multicolumn{1}{l|}{}                                                & \textbf{Uncertain}     & 15                                 & 34                                     & 3                                       \\ \hline
\multicolumn{1}{l}{\textbf{Total}}                                           & \multicolumn{1}{l}{}  & \multicolumn{1}{c}{\textbf{181}}           & \multicolumn{1}{c}{\textbf{299}}               & \multicolumn{1}{c}{\textbf{15}}                 \\ \hline
\end{tabular}
\end{table}
%\begin{figure*}[htbp]
%  \centering
%  \includegraphics[width=\textwidth]{eps/12TailedCase}
%  \caption{Relación entre los tipos de cola y la significación estadística de los tests}
%  \label{f\textbf{ig:12TailedCase}}
%\end{figure*}

\subsection{Review findings}\label{sec:paper_review:finding}

For our perspective, the main findings of the review are:

\begin{itemize}

\item \textbf{Failure to control for bias}: 

  \begin{enumerate}

  \item Are hypotheses explicitly defined? 
  \textbf{No}. Only 55\% of the papers contain a research hypothesis. However, it is doubtful that the lack of definition is associated with HARKing. When the research hypotheses are not present, the authors provide research questions. Hypotheses may or may not define the type of tails. The current situation can be explained more easily by inconsistent reporting and/or limited statistical expertise than HARKing.

  \item Are there \textit{post hoc} tests?
  \textbf{Yes, but in lower rates than expected}. Only 31\% of the papers contain \textit{post hoc} tests. Most of these \textit{post hoc} tests (80\%) are unnecessary because the experiment already has produced statistically significant results.

  \end{enumerate}
   
\item \textbf{Analysis flexibility}:
   
  \begin{enumerate}\setcounter{enumi}{2}

  \item Which analysis procedures do ESEM experiments apply?
  \textbf{Regular procedures}, such as t-test, ANOVA or their non-parametric counterparts. Sophisticated tests are not used.

  \end{enumerate}
   
\item \textbf{Data dredging}:
   
  \begin{enumerate}\setcounter{enumi}{3}

  \item How many explicit hypotheses are defined per experiment?
	\textbf{2.41 hypotheses on average}. Most of the experiments declare $\leq 2$ hypotheses.

  \item How many tests are conducted per experiment?
	\textbf{7.95 tests in average}. There are more tests than hypotheses due to the existence of independent variables not mentioned in the hypotheses.

%  \item How many tests are conducted per hypothesis?
%
%\odnote{explicar} Max:44, Min:0, Avg: 3.08

  \item What is the ratio of statistically significant tests?
	\textbf{0.37}. This value is higher than the ratio reported by J\o rgensen et al. \cite{jorgensen2016incorrect} (51\%).

  \item Does the hypothesis tail match the analysis tail?
  \textbf{Yes, in general}. There are some inconsistencies and errors, but no evidence of bias. Most of the tests lack an explicit definition of the type of tail. This problem seems to be connected to the confusion around research/statistical hypotheses and tails.
   
  \item Are 1-tailed tests used to increase power?
  The researchers' intentions cannot be inferred from the data. However, the high number of statistically significant results associated to 1-tailed tests \textbf{suggests a positive answer}.

  \end{enumerate}

\end{itemize}

\section{Survey to SE experimenters}\label{sec:survey-experimenters}

The literature review reported in the previous section shows some issues in the conduction of experimental research in ESEM, e.g., the inconsistent use of research/statistical hypotheses. RQ2 aims to clarify these issues. We performed a survey with SE experimenters, inquiring how they plan an experimental study and their beliefs about the associated concepts, e.g., the type of tails. The sections below report the survey design, execution, and results.

\subsection{Survey design}\label{sec:design_survey_researchers}

The survey design is based on Kitchenham et al. \cite{kitchenham2008personal} and Punter et al.'s \cite{punter2003conducting} guidelines. The questions try to clarify why researchers perform some practices that surfaced during the literature review, in particular:

\begin{enumerate}

\item The lack of clear relationships between research and statistical hypotheses, and the associated tails.

\item The seemingly arbitrary choice of 1-tailed and 2-tailed tests.

\item The errors and inconsistencies in statistical tests.

\end{enumerate}

The first version of the survey consisted of 21 mandatory questions and six optional questions. Several questions address the same topic to avoid misinterpretations; this makes the survey more time consuming than we initially wanted. The respondent is allowed to add comments or opinions using free text. Five mandatory questions were removed, and eight new ones (mandatory and optional) were created after a thorough review by O. Dieste and N. Juristo. In a second stage, two independent researchers (M. Solari y O. G\'omez) piloted the survey; their feedback improved the text of 4 optional questions. The final version of the survey is available at \url{https://goo.gl/QS1ati}.

The population is defined as any SE experimenter. The sample was collected as follows: We collected the emails of the experimenters who published experiments in the most representative conferences and journals of the experimental Software Engineering community\footnote{Retrospectively, we believe that the Information and Software Technology journal (IST) had to be included as well. Nevertheless, notice that the authors that publish at IST and the other four outlets overlap considerably.} between 2012 and 2015, i.e., ESEM, the International Journal on Empirical Software Engineering (EMSE), the IEEE Transactions on Software Engineering (TSE), and the International Conference on Software Engineering (ICSE). We obtained a total of 403 authors' emails (95 of ESEM, 93 of EMSE, 124 of TSE, and 91 of ICSE). When the same author appears in several outlets, she/he is only considered once (in the following order: ESEM, EMSE, TSE, ICSE).

\subsection{Execution}

Respondents were contacted by email and periodically reminded. The data were collected in approximately 3 months. Each group of authors (ESEM, EMSE, TSE, ICSE) responded to the survey at a different time, so the provenance of each respondent could be identified (in all other respects, the identity of the respondent was not known to the researchers).

The experimenters completed the survey in 8-12 minutes (depending on the feedback provided). A survey of this duration is unlikely to produce fatigue, to the point of representing a threat to validity \cite{kitchenham2008personal}. 45 researchers answered the questionnaire (29 in full, 16 partially). The response ratio was 11.20\% (45/403), which is comparable with the response ratios achieved by other SE surveys, e.g., \cite{de2014survey}. The data obtained are available at \url{https://goo.gl/1X3Px9}.

\subsection{Results}\label{sec:results_and_interpretation-survey-experimenters}

The survey results are summarized in trees and double-entry tables, as in the previous section. The respondents' comments (submitted as free text) have been coded to highlight the underlying themes.

\subsubsection{Relationship between the research hypothesis, statistical hypothesis, and tails}

The vast majority of researchers (86\%) claim that they include research hypotheses in their reports, as shown in Figure \ref{fig:HInvHestResSurveyExperimenters}. This number is 30 points higher than Fig.~\ref{fig:RQ11}, where only 55\% of the experiments contain research hypotheses. The same pattern (30\% difference) appears in the statistical hypotheses. 59\% of researchers say that they always include statistical hypotheses in their investigations, while Fig.~\ref{fig:RQ11} indicates that this only occurs in 29\% of the cases.

\begin{figure*}[htbp]
  \centering
  \includegraphics[width=0.65\textwidth]{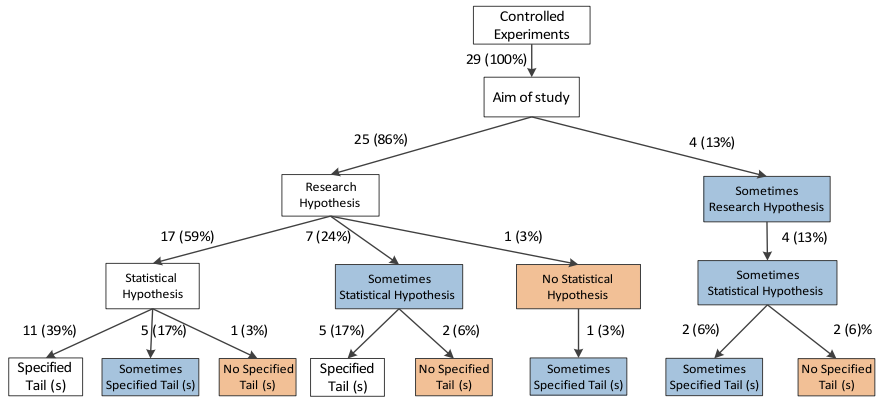}
  \caption{Relationship between research hypotheses and statistical hypotheses}
  \label{fig:HInvHestResSurveyExperimenters}
\end{figure*}

It could be argued that the articles published in ESEM are particularly defective. Fig.~\ref{fig:RQ21} shows the equivalent to Fig.~\ref{fig:RQ11}, but restricting the responses to the experimenters who have published in ESEM. The experimenters argue that they include research/statistical hypotheses in 80\% and 60\% of the cases, respectively.

\begin{figure*}[htbp]
\centering
\includegraphics[scale=0.25]{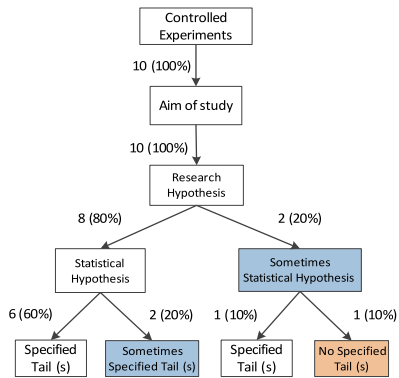}
\caption{Relationships between aims, research hypotheses, statistical hypotheses, and tails} 
\label{fig:RQ21}
\end{figure*}

Figure \ref{fig:HInvHestResSurveyExperimenters} shows that researchers know the hypothetical-deductive method. Only a minority of researchers (3\%) state that they do not use statistical hypotheses, or specify the tails of alternative hypotheses (17\%). It stands in stark contrast with the literature review results, where the figures are the opposite (71\% and 51\% respectively). 

%Finally, as shown in the Table~\ref{table:Clear-connection-or-relationship-between}, a substantial number of researchers (31\%) consider that the relationship between research hypotheses and statistical hypotheses is less direct from what the commonly accepted theory indicates.
%
%\input{tex-files/Clear-connection-or-relationship-between}
%
When a participant provided a response apparently in opposition to the usual practice (e.g., she/he does not define statistical hypotheses), the survey asked for the reasons why. 
%Figure \ref{fig:comments} shows the answers about it 
After a process of refinement and classification (available at \url{https://goo.gl/FTxnV5}), we obtained the following advice from experimenters:

%\begin{figure*}[htbp]
%\centering
%\includegraphics[width=\textheight, angle=90]{dotgraph/relationships.eps}
%\caption{Summary of experimenters' comments on research \& statistical hypotheses}
%\label{fig:comments}
%\end{figure*}

\begin{itemize}

\item \textbf{Research hypothesis:} There are two noteworthy aspects: a) A significant number of researchers believe that the research hypothesis can be easily inferred, and b) \textit{exploratory studies}\footnote{Experimenters differentiate between exploratory and confirmatory studies. The former seek relationships between variables, whereas the latter try to confirm/reject given research/statistical hypotheses.} do not need research hypotheses. 

\item \textbf{Statistical hypothesis}: The reasons given are essentially the same as above: a) they are obvious, or b) they are not useful in exploratory studies. A third reason is that a decision about which hypotheses to test cannot be taken during the experimental design and it should be deferred until the experiment has been executed. This goes against the recommended practice, although a more likely interpretation is that the respondent is referring to conducting exploratory studies.

\item \textbf{Tails}: The situation is, again, quite similar: a) the tails are evident or can be inferred, or b) they are useless since they can not be defined during the design (that is, in exploratory studies). The researchers show a preference for the use of two tails, which could be related, again, with the exploratory studies, since the lack of the direction of the effect is, precisely, one of their main characteristics.
%
%\item \textbf{Relationship between research hypothesis and statistical hypothesis}: Some researchers have disagreed about this relationship. In all cases, the discrepancy refers to the difficulty of translating the research questions into statistical hypotheses. The reasons for this difficulty are not mentioned. 

 \end{itemize}

\subsubsection{Choice of 1-tailed or 2-tailed tests}

%We asked the researchers when they believe that the 1/2-tailed tests are applicable. Most researchers believe that 2-tailed tests are used in exploratory studies, while 1-tailed tests are more useful in confirmatory studies. However, in the latter case, a large number of researchers also suggest the use of the 2-tailed test (see Table \ref{table:TypeStudyby12tailedALL}). 

In the previous section, we have reported that experimenters exhibit a preference for 2-tailed tests. The Tables \ref{situations-used-1-tailed} and \ref{situations-used-2-tailed} explain the reason why. In 23\% of cases, researchers indicate that 2-tailed tests should always, or in most situations, be used. 1-tail tests are generally discouraged. In 14\% of the cases, experimenters believe that they should never be used.

\begin{table}[h!]
\scriptsize
\centering
\caption{Situations in which 1-tailed test should be used}
\label{situations-used-1-tailed}
\begin{tabular}{p{5cm} c c }
\hline \hline
\textbf{When 1-tailed tests should be used?}                                                & \textbf{Frequency} & \textbf{Questionable?} \\ \hline
-                                                                                           & 1                   &                     \\ 
When the effect has a clear direction                                                  & 3                   &                     \\ 
When the researcher is only interested in certain effect direction                  & 3                   &                     \\ 
The effect can only occur in a certain direction, according to prior knowledge     & 4                   &                     \\ 
It depends on the existing knowledge about a phenomenon                                        & 1                   &                     \\ 
The effect can only occur in one direction                                               & 1                   &                  \\ 
The hypothesis is directional                                                                    & 9                   &                     \\ 
The distributions are asymmetrical                                                          & 1                   & Probably                  \\ 
Never in SE                                                                                 & 3                   &                     \\ 
Never, unless there is a good reason to assume an effect in a certain direction & 1                   &                     \\ 
Post-hoc tests                                                                              & 1                   & Yes                  \\
Test the alternative hypothesis                                                             & 1                   & Maybe                  \\\hline
\textbf{Total}                                                                         & \textbf{29}         &                     \\ \hline \hline
\end{tabular}

\end{table}

\begin{table}[]
\scriptsize
\centering
\caption{Situations in which 2-tailed test should be used}
\label{situations-used-2-tailed}
\begin{tabular}{p{5cm} c c }
\hline \hline
\textbf{When 2-tailed tests should be used?}                                              & \textbf{Frequency} & \textbf{Questionable?} \\ \hline
-                                                                                         & 3                   &                     \\ 
Any difference is of interest                                                        & 3                   &                     \\ 
The statistical power is enough                                                        & 1                   & Yes                  \\ 
In most instances                                                          & 5                   &                     \\ 
There is more than one independent variable                                                  & 1                   & Yes                  \\ 
The statistical distribution has two tails                                               & 1                   & Yes                  \\ 
The null hypothesis suggests that there is no difference between two distributions                & 2                   & Yes                  \\ 
No, there should be prior knowledge that indicates the direction of the effect            & 1                   &                     \\ 
The effect direction is unknown                                                      & 11                  &                     \\ 
Always, since in SE it is unlikely that the effects have a well-defined direction & 2                   &                     \\ \hline
\textbf{Total}                                                                       & \textbf{29}         &                     \\\hline \hline
\end{tabular}
\end{table}

The reasons given for the choice of 1-tailed or 2-tailed tests are completely reasonable. In 72\% of cases, the researchers indicate that the 1-tailed tests should be used when: a) the direction of the effect is known, or b) the experimenter is only interested in one of the directions of the effect. For the 2-tailed tests, the situation is similar; in 47\% of cases, the reason given is the uncertainty of the effect's direction. 

A remarkable aspect in Tables \ref{situations-used-1-tailed} and \ref{situations-used-2-tailed} is that some responses are questionable, perhaps pointing at issues with experimental statistics.

\subsubsection{Usage of statistical tests}

We have tried to find out to what extent the researchers correctly handle the concepts of 1/2 tails at the level of statistical tests. Hence, we have made 3 specific questions:
 
\begin{table}[!htbp]
\scriptsize
\centering
\caption{Which statistical tests do you use 1-tailed tests or 2-tailed test with?}
\label{table:StatisticalTestwith12Tailed}
\begin{tabular}{
>{\centering\arraybackslash}p{2cm}
>{\centering\arraybackslash}p{0.5cm}
>{\centering\arraybackslash}p{0.5cm}
}
\cline{2-3}
                                                                  & \multicolumn{1}{c}{\textbf{1-tailed}} & \multicolumn{1}{c}{\textbf{2-tailed}} \\ 
\hline \hline
\multicolumn{1}{l}{Student's t-test (paired/unpaired)}         & \multicolumn{1}{c}{62\%}      & \multicolumn{1}{c}{62\%}  \\ 
\multicolumn{1}{l}{Mann-Whitney / Wilcoxon (paired/unpaired)} & \multicolumn{1}{c}{59\%}     & \multicolumn{1}{c}{79\%}      \\ 
\multicolumn{1}{l}{Chi-squared test}               & \multicolumn{1}{c}{\cellcolor[HTML]{EFEFEF}\textbf{31\%}}       & \multicolumn{1}{c}{59\%}      \\ 
\multicolumn{1}{l}{ANOVA}                                       & \multicolumn{1}{c}{\cellcolor[HTML]{EFEFEF}\textbf{28\%}}     & \multicolumn{1}{c}{66\%}          \\ 
\multicolumn{1}{l}{ANOVA With Repeated Measures}                & \multicolumn{1}{c}{\cellcolor[HTML]{EFEFEF}\textbf{10\%}}  & \multicolumn{1}{c}{38\%} \\ 
\multicolumn{1}{l}{Kruskal-Wallis test}                         & \multicolumn{1}{c}{\cellcolor[HTML]{EFEFEF} \textbf{41\%}}     & \multicolumn{1}{c}{55\%}      \\ 
\multicolumn{1}{l}{Linear regression}                           & \multicolumn{1}{c}{\cellcolor[HTML]{EFEFEF} \textbf{3\%}}   & \multicolumn{1}{c}{31\%}     \\ \hline \hline  
\end{tabular}
\end{table}

\begin{itemize}

\item \textbf{Which tails can be used with which statistical tests?} The responses (experimenters could select both the 1-tail and 2-tail options, so the percentages do not add up to 100\%) are shown in Table \ref{table:StatisticalTestwith12Tailed}. Some answers were incorrect (shaded cells in the table). 31\% of experimenters say they use one tail with the ANOVA, and 41\% with the Kruskal-Wallis tests when both are inherently 2-tailed. A similar figure (31\%) is obtained for the Chi-squared test, which is almost universally used for contingency table analysis, also inherently 2-tailed. 

It is highly unlikely that the experimenters are thinking about the underlying statistic. For instance, F-test is used to assess the statistical significance in the case of the ANOVA. The F-test is 1-tailed, but the ANOVA is 2-tailed. However, the percentage of responses in the 2-tailed column for ANOVA is higher (and also correct) than the percentage of the 1-tailed column; in our opinion, the answers in the 1-tailed column are mistakes.

\item \textbf{Have you tried several statistical techniques during the analysis?:} 7\% of researchers say they do it. 52\% of researchers say they have done so, at least on some occasions.   

\item \textbf{Have you ever made a change in the type of tail during the analysis?} The vast majority of the experimenters never switched 1-tailed tests (from $\leq$ to $\geq$, and \textit{vice versa}) or 2-tailed tests into 1-tailed ones. However, a sizable number of experimenters (21\%) switched from 2-tailed to 1-tailed tests, as shown in Table \ref{table:Change2-tailedInto1-tailed2}. When this occurs, significant results were obtained in 10.5\% of cases. 

\end{itemize}

%\input{tex-files/Direction-1tailed-OtherDirection-1tailed}
% \usepackage{multirow}
\begin{table}[H]
\scriptsize
\centering
\caption{Did you ever change a 2-tailed hypothesis into a 1-tailed one?}
\label{table:Change2-tailedInto1-tailed2}
\begin{tabular}{p{2.1cm}
>{\centering\arraybackslash}p{0.3cm}
>{\centering\arraybackslash}p{0.3cm}|
>{\centering\arraybackslash}p{1.8cm}
>{\centering\arraybackslash}p{1.8cm}}
\cline{4-5}
                                                                                                                &                                            &                        & \multicolumn{2}{p{4cm}}{\textbf{When you switched the 2-tailed hypothesis into a 1-tailed one, did it turn out that a statistically significant result came to light?}} \\ \hline \hline
\multicolumn{1}{l}{}                                                                                          & \multicolumn{1}{l}{}                      &                        & Yes                                                                 & \cellcolor[HTML]{EFEFEF}10.5\%                                                                \\ \cline{4-5} 
\multicolumn{1}{l}{}                                                                                          & \multicolumn{1}{l}{\multirow{-2}{*}{Yes}} & \multirow{-2}{*}{21\%} & No                                                                  & 10.5\%                                                                                        \\ \cline{2-5} 
\multicolumn{1}{p{2cm}}{\multirow{-3}{2.3cm}{\textbf{Did you ever change a 2-tailed hypothesis into a 1-tailed one?}}} & \multicolumn{1}{l}{No}                    & 79\%                   & \multicolumn{2}{l}{-}                                                                                                                       \\ \hline \hline
\end{tabular}
\end{table}

\subsubsection{Survey findings}

In our opinion, the main findings of the survey are:

\begin{enumerate}

\item Relationships between research, statistical hypotheses, and tails: 
Theoretically speaking, \textbf{researchers know how to use the statistical concepts}. However, they \textbf{do not apply them in practice}. The reason is that experiments do not seem these concepts (research hypothesis, tails) useful to \textbf{conduct exploratory research}.

\item Choice of 1-tailed and 2-tailed tests: 
Same as before, \textbf{experimenters know when to use 1-tailed or 2-tailed tests}. However, some experimenters \textbf{make mistakes when proposing usage scenarios}.

\item Usage of statistical tests: 
A sizable number of experimenters \textbf{make mistakes in the types of tails associated with some tests}. In some cases, \textbf{tails are switched}, with the likely intention of increasing the power of the test.

\end{enumerate}
\section {Threats to validity}\label{sec:threats-validity}

The threats to validity will be reported following Yin \cite{yin2013case} and Creswell \cite{creswell2002-research}. Both works, compared to Shadish et al. \cite{Shadish2002}, take into consideration qualitative studies (as this one). According to those authors, threats to validity are classified into four types: internal, external, construct and reliability.

\textbf{Internal validity:} It refers to the inferences made on data. The existence of unknown variables may influence the results of the statistical tests or other analysis procedures \cite{creswell2002-research}. This threat can operate both in the literature review and the survey. To mitigate this threat:

\begin{itemize}

\item The literature review was conducted by three researchers. All relevant steps (paper screening, selection, data extraction, and analysis) were performed collaboratively. We set controls, e.g., double-check of the data extraction forms, re-calculation of tables, etc. to guarantee the quality of the data.

\item The survey was piloted and evaluated by independent researchers. The number and complexity of the questions were limited to avoid respondent's fatigue (the survey could be filled out completely in less than 15 minutes). To improve the veracity/accuracy of the responses, participation was voluntary, and anonymity was secured.

\end{itemize}

\textbf{External validity:} This threat appears when we generalize the results beyond the context in which they were obtained. The risks are different for the review and the survey:

\begin{itemize}

\item The literature review was specifically targeted to ESEM experiments and, consequently, the ''community'' behind these studies. We do not aim to generalize to other communities; in fact, the specific analysis of the ESEM community is one of the goals of this paper. Further research, e.g., targeting other venues where experimental papers are regularly published, would be necessary to assess the situation of the general SE community.

\item The survey has a more general character. For that reason, the sample was not restricted to ESEM researchers. We also include researchers that published experimental papers at EMSE, ICSE, and TSE. We believe that the researchers that publish in these outlets are representative of the general population of experimental researchers.

\end{itemize}

\textbf{Construct validity:} This threat operates when the study uses variables and metrics that do not represent the underlying theoretical constructs accurately. We have addressed this threat conducting previous research on (1) statistical errors in SE \cite{Reyes2018icse} and (2) experimental problems in the sciences (not published yet). These previous studies allowed us to design rigorous instruments (data collection forms, questionnaire); these instruments were also double-checked and/or piloted.

\textbf{Reliability:} It refers to how trustable and repeatable the study results are. To mitigate this threat, we have documented (as much as possible) all methodological steps, e.g., study screening and primary study selection. Documents are available in Google Drive\texttrademark; the URLs are have been provided throughout the manuscript. Likewise, the raw data and analysis results are publicly available in the same form.
\section{Discussion and conclusions}\label{sec:conclusion}  

We cannot provide an authoritative answer to \textbf{RQ1: Is there evidence of publication bias at ESEM?} The signs are ambiguous. On the one side:

\begin{itemize}

\item Hypotheses and tails are left undefined frequently.

\item A large percentage of papers perform \textit{post-hoc} tests.

\item Multiple testing is the norm. The number of tests conducted on average per experiment is 7.95.

\item 1-tailed tests are associated with higher significant test ratios.

\end{itemize}

On the other side, these practices do not seem oriented to achieve statistically significant results that underlie publication bias:

\begin{itemize}

\item Most (80\%) of the \textit{post-hoc} tests were unnecessary because the experiment yielded significant results already. 

\item The ratio of statistically significant tests is 0.37, quite close to the average power of SE experiments, as reported by J\o rgensen et al. \cite{jorgensen2016incorrect}.

\end{itemize}

In our opinion, \textbf{publication bias is not strongly influencing the research agenda, not at least in ESEM}. Experimenters perform some questionable practices: (1) lack of definition of the experimental hypotheses and (2) multiple testing. Both practices probably increase the significant test ratio (0.3 is likely higher than the actual average power of SE experiments) and consequently make some (unintended) pressure on authors, reviewers, and editors. 

However, such increment seems to be a collateral effect of how SE experiments are being conducted, but not a major driver. \textbf{RQ2: Why do researchers carry out questionable practices?} aimed to find out why researchers perform those questionable practices. The most frequent answer was that they perform \textbf{exploratory research}. It is generally agreed that the SE discipline lacks sound scientific knowledge. It is perfectly reasonable, in our opinion, that experimenters perform ''reconnaissance'' studies, without clear \textit{a priori} hypotheses and using multiple testing to find relationships among variables. In turn, we should adapt our reporting procedures (among other potentially useful measures, such as pre-registration) to acknowledge the exploratory character of the studies and properly qualify the strength of the evidence provided in the experiment.

Finally, we also have observed some weaknesses in the experimenters' statistical knowledge. This problem was pointed out by other researchers, e.g., \cite{Vegas-2016-Crossover-Designs-ESE,Kitchenham2019ease} previously. The ESEM community (and the overall SE community, as well) should establish measures to improve experimenters' statistical skills. 

\section{Acknowledgments}
We wish to express our appreciation to the experimenters that participated in the survey. This work has been partially supported by the Spanish Ministry of Economy and Competitiveness research grant TIN2014-60490-P, the '' Laboratorio Industrial en Ingenier\'ia del Software Emp\'irica (LI2SE)'', ESPE research project.

%Esto evita que las tablas penetren dentro de la bibliografía
\FloatBarrier

\bibliographystyle{ACM-Reference-Format}

\bibliography{paper}

\end{document}